\documentclass[showpacs,preprintnumbers,twocolumn,amsmath,amssymb]{revtex4}
\usepackage{graphicx}
\usepackage{dcolumn}
\usepackage{bm}

\begin{document}

\title{Andreev reflection and tunneling spectrum in a superlattice of  metal-superconductor junctions}
\author{W. LiMing}\email{wliming@scnu.edu.cn}
\author{Jiayun Luo, Xiaoxue Cai, Ke Sha, Liangbin Hu}
\affiliation{Dept. of Physics, and Laboratory of Quantum Information Technology, School
of Physics and Telecommunication Engineering, South China Normal University, Guangzhou
510006, China}
\begin{abstract}The tunneling spectrum of an electron and a hole in a superlattice of NS junctions is computed using the BTK approach and the transfer matrix method. It shows sharp resonances at some energies above the superconducting gap.  The sharper the resonance is the more layers the superlattice has. We find  for the first time a  mechanism to balance the incident and outgoing currents on the superlattice  by averaging over the phase between the incident electron and the incident hole. This mechanism is more natural and physical than those in literatures.
\end{abstract}
\date{\today}
\maketitle 
\section{Introduction}
The Andreev reflection in normal metal-superconductor(NS) junctions has  attracted great interests of researchers in the field of superconductivity in recent years\cite{Blonder,Satoshi, Yang, wliming}. The tunneling spectrum of electrons on a NS junction is sensitive to the superconducting gap of the superconductor(SC), thus providing an important technique to measure the gaps and gap properties of SCs. When an electron tunnels into a SC from a metal there appears a hole reflection to conserve the electric charge, which is called the Andreev reflection, since the incident electron may form a Cooper pair in the SC. A commonly used theoretical method to investigate the Andreev reflection is the Blonder-Tinkham-Klapwijk(BTK) approach\cite{Blonder}, which takes the interface of a NS junction as a $\delta(x)$ potential barrier. This theory has been widely and successfully applied to systems like NS junctions\cite{Tanaka}, ferromagnet-SC junctions\cite{Dong2}, etc.  Wei, Dong and Xing {\it et al} studied the tunneling spectrum in metal-SC-metal(NSN) junctions using the BTK approach\cite{Wei,Dong}. They found that the incident and outgoing currents in both sides do not balance each other\cite{Wei}. Thus they claimed earlier that the BTK approach were not suitable for the Andreev reflection in NSN junctions and treated them in the Landauer-Buttiker formalism\cite{Buttiker}. Later they proposed a mechanism to balance the currents by adjusting the chemical potential of the SC inside the NSN junctions\cite{Dong}. This mechanism seems reasonable since the unbalance of currents will change the charge density and thus the Fermi surface of the SC. In a similar idea we proposed recently another mechanism that the interface of the NS junctions is charged by the unbalanced currents\cite{wliming}. 

Although these mechanisms successfully balanced the incident current and the outgoing current, however, they mis-considered the phase between the incident electron and hole. We found that different phases just balance the incident current and the ougoing current. After averaging over the phases the incident and outgoing currents coincide each other through the whole range of bias voltage. The charge density, the Fermi surface, and the interface charge state won't change at all. This provides us a more natural and physical mechanism for the theory of multiple NS junctions. In the present paper we study the superlattice of multiple NS junctions using this new mechanism.

\section{formalism}
 A NS junction superlattice is shown in Fig.1, where metals and SCs with width $L_N$ and $L_S$ are sandwiched periodically but the outmost layers are metals working as two leads.  A thin insulating interface exists in every junction to be treated as a $\delta(x)$ potential barrier in the BTK approach. An electron and a hole are incident into the superlattice under a bias voltage on both sides. There is no incident electron in the right metal, thus $b_{n+1}=0$ and in the left metal no incident hole $d_1=0$. Due to the potential barriers the electron and the hole in one layer are partially reflected and partially transmitted into the next layer. In the SC only a pseudo-particle can propagate above the energy gap $\Delta$. When the bias voltage is smaller than $\Delta$ only Cooper pairs can do in the SC. This causes a hole reflected from the interface, which is called the Andreev reflection in literatures.

\begin{figure}
\includegraphics[width=6cm,height=4cm]{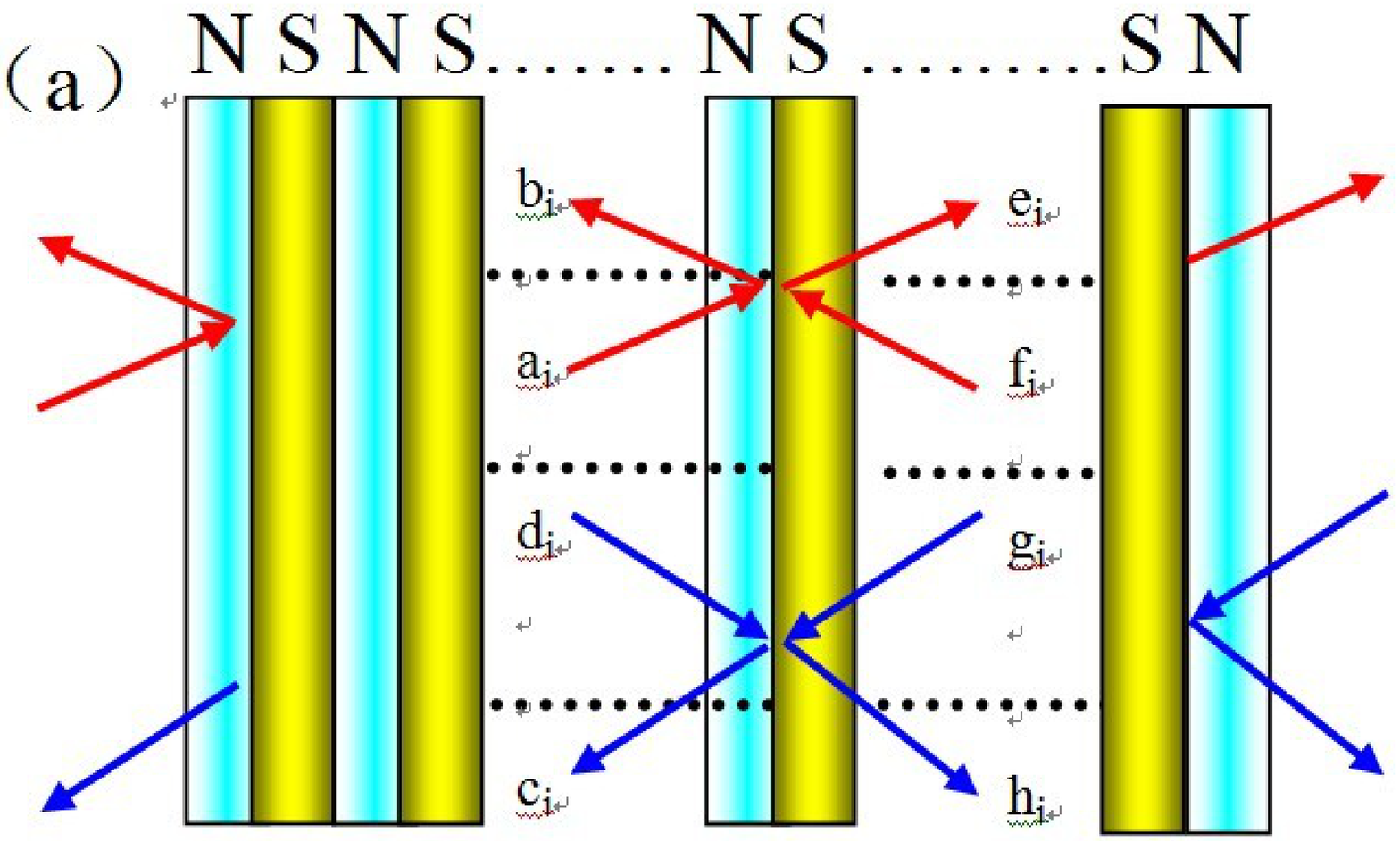}
\includegraphics[width=8cm,height=5cm]{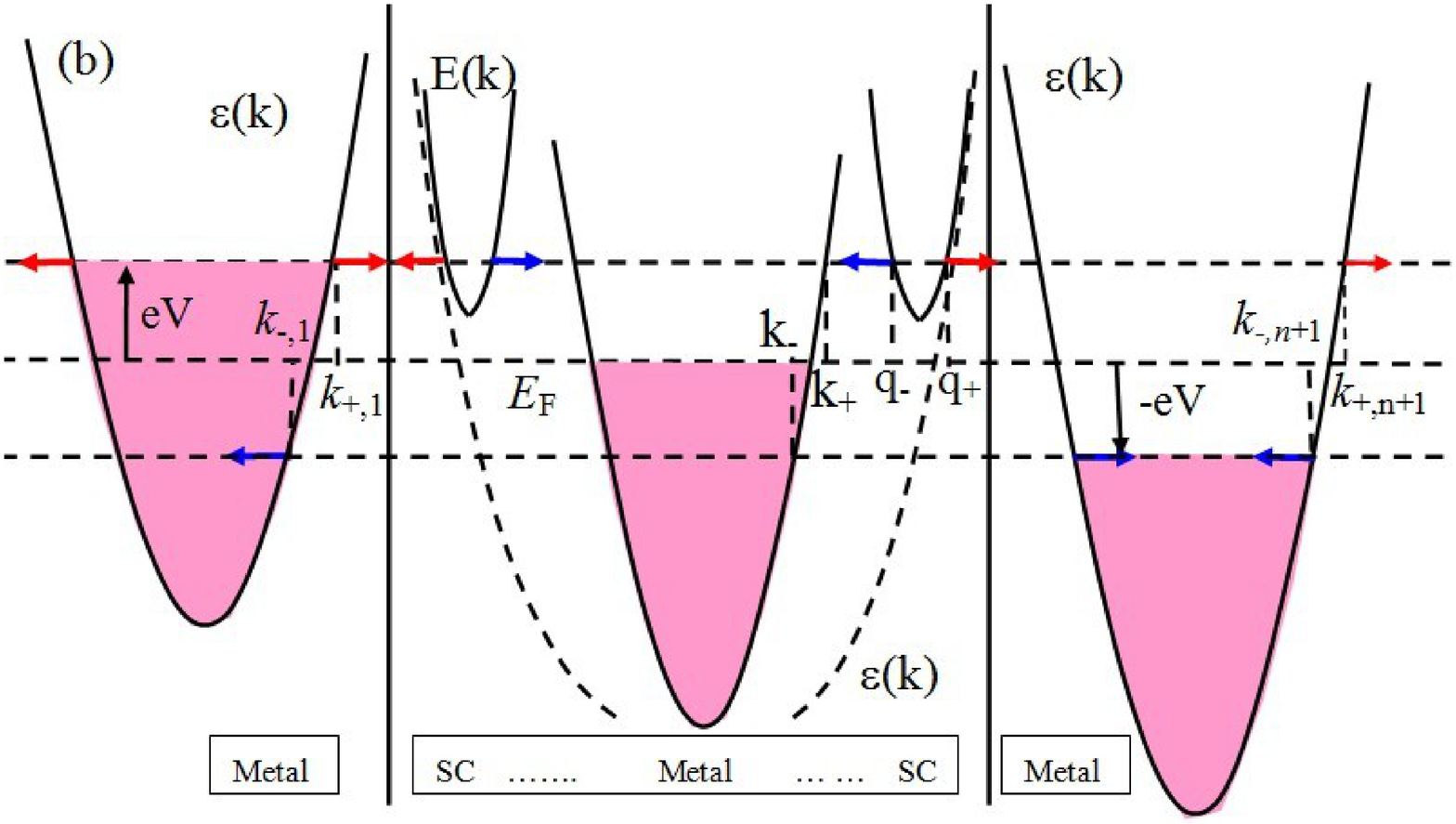}
\caption{(a) a NS junction superlattice with an incident electron and an incident hole under a bias voltage. The metal  and the SC are colored by blue and yellow, respectively. An electron is incident from the left metal and partially tunnels into the adjacent SC and is partially reflected. In the right side there is no incident electron, $b_{n+1}=0$ and in the left side $d_1=0$. (b) The fermi surfaces and energy bands of metals and superconductors. Except for the two metal leads the metals and SCs have 0 bias voltage. }
\end{figure}

The tunneling process of an electron and a hole in the NS junction superlattice is described by the Bogoliubov-de Gennes (BdG) equation\cite{BdG}
\begin{align}
i\hbar {\partial \over \partial t} \begin{pmatrix} u({\bf r},t)\\v({\bf r},t)\end{pmatrix} = \begin{pmatrix} H_0 &\Delta \\ \Delta^* & -H_0\end{pmatrix} \begin{pmatrix} u({\bf r},t)\\v({\bf r},t)\end{pmatrix}\\
H_0 = -{\hbar^2\over 2m} \nabla^2 + V({\bf r}) -\mu
\end{align}
where $V({\bf r})= U\delta(x)$ on the interfaces.

When the bias voltage satisfies $eV > \Delta$ an incident particle in metals tunnels into the SC through the interfaces and propagate as a quasi-particle inside the SC. It is then partially reflected by the following interface and partially tunnels into the following metal.  The wave functions of the electron and the hole inside the $i$th metal layer and the $i$th SC layer are written as
\begin{align}
\psi^i_{N}= \left(
  \begin{array}{c}  a_i e^{ik_+ x}+b_i e^{-ik_+ x}\\   c_i e^{i k_- x}+ d_i e^{-i k_- x}\\
  \end{array}\right) \label{wf1}
\\
\psi^i_{SC}=\biggl[(e_i e^{iq_+ x}+ f_i e^{-iq_+ x})
\left(
  \begin{array}{c}   u_{+}\\  v_{+}\\  \end{array} \right)   \nonumber\\
  + (g_i e^{iq_- x} + h_i e^{-iq_- x})
\left(   \begin{array}{c}     u_{-}\\    v_{-}\\   \end{array} \right) \biggr]
\label{wf2}
\end{align}
where $\left(
  \begin{array}{c}   u_{+}\\  v_{+}\\  \end{array} \right)$ and $\left(
  \begin{array}{c}   u_{-}\\  v_{-}\\  \end{array} \right)$ are the quasi-eletron ($\epsilon_q>0$) and quasi-hole ($\epsilon_q<0$) wave functions with the same energy $E$ in the SC layers, respectively. They are given by
\begin{align}
u = \sqrt{{1\over 2} + {\epsilon_{q} \over 2E}}, v = {|\Delta|\over \Delta}\sqrt{{1\over 2} - {\epsilon_{q}\over 2E}}\\
\epsilon_q ={\hbar^2 q^2\over {2m}}-\mu=\pm \sqrt{E^2 - |\Delta|^2}
\end{align}
The wave vectors are determined by
 \begin{align}
 E\equiv eV=eV+{\hbar^2 k_{+,1}^2\over {2m}} - \mu =  \mu - {\hbar^2 k_{-,1}^2 \over {2m}}-eV\\
 \quad k_{+,1}=k_F = \sqrt{2m\mu/\hbar^2}\\ k_{-,1}/k_F=\sqrt{1- {2E/\mu}}\\
\quad k_{\pm, i}/k_F = \sqrt{1\pm {E/\mu}}, 1<i<n+1\\
\quad q_\pm/k_F = \sqrt{1\pm \sqrt{E^2-\Delta_q^2}/\mu}\\
2eV={\hbar^2 k_{+,n+1}^2\over {2m}} - \mu ,\quad 0 =  \mu - {\hbar^2 k_{-,n+1}^2 \over {2m}}\\
\quad k_{+,n+1}/k_F=\sqrt{1+{2E/\mu}}, \quad k_{-,n+1}= k_F
\end{align}
When $eV<\Delta$, $ q_\pm$ become complex numbers, so that Cooper pairs appear and the wave functions in the SC become damping traveling waves.
In this case one has
 \begin{align}
\epsilon_q  = \pm i\sqrt{ |\Delta|^2- E^2}\\
\begin{pmatrix}u\\v\end{pmatrix}_\pm ={1\over \sqrt 2}\begin{pmatrix}1 \\{E- \epsilon_q\over \Delta}\end{pmatrix}
\end{align}
The coefficients $a,b,c,d,e,f,g,h$ in (\ref{wf1}) to (\ref{wf2}) are obtained from the boundary conditions of wave functions on the interfaces, where the wave functions are continuous but due to the $\delta(x)$ barriers
the first derivatives of them are not, i.e.,
\begin{align}
\psi_{1}(X_i)=\psi_{2}(X_i)\\
\psi'_{1}(X_i)-\psi'_{2}(X_i)+Zk_F\psi_{1}(X_i)=0
\end{align}
where 1 and 2 denote the adjacent metal and SC layers, and $Z=\frac{2mU}{\hbar^{2}k_F}$ and $X_i$ is the position of the interface. These boundary conditions give the following matrix equations
\begin{align}
(a_i, b_i, c_i, d_i)^T = M_1(X_i)(e_i,f_i,g_i,h_i)^T\\
(e_i,f_i,g_i,h_i)^T=M_2(Y_{i}) (a_{i+1}, b_{i+1}, c_{i+1}, d_{i+1})^T
\end{align}
for $i = 1,2, ..., n$, where $M_1(X_i)$ and $M_2(Y_{i})$ are called the transfer matrices at interface $X_i$ between the metal and SC layers and interface $Y_i$ between the SC and metal layers. Using this matrices we find the relation between the coefficients of the incident and outgoing particles in the following form
\begin{align}
\begin{pmatrix}a_1\\ b_1\\c_1\\ d_1\end{pmatrix} = \prod_i M_1(X_i) M_2(Y_i) \begin{pmatrix}a_{n+1}\\ b_{n+1}\\ c_{n+1}\\ d_{n+1}\end{pmatrix}
\end{align}
Assigning $a_1=1, d_1=0, b_{n+1}=0$, and $c_{n+1}=exp(i\phi)$ for an incident electron and an incident hole with a phase difference $\phi$,  other coefficients can be obtained relative to $\phi$ from this matrix equation. These coefficients determine the the incident and outgoing current densities and the differential electric conductance\cite{wliming}.
\begin{align}
J_1 \sim[k_+(1-|b_1(eV)|^2)+k_- |c_1(eV)|^2]\\
J_{n+1} \sim[k_-(1-|d_{n+1}(eV)|^2)+k_+|a_{n+1}(eV)|^2]\\
{dI_1(eV)\over dV}\sim 1-|b_1(eV)|^2+|c_1(eV)|^2\\
{dI_2(eV)\over dV}\sim 1-|d_{n+1}(eV)|^2+|a_{n+1}(eV)|^2
\end{align}

In general the incident current and the outgoing current do not match each other in a NS junction superlattice even in NSN junctions as pointed by Xing et al\cite{Dong}. The key idea of this work is that the currents depend on the phase difference $\phi$. After averaging over the phase the incident current just balances the outgoing current and so does the differential conductance.

\section{Results of computation}
The phase difference $\phi$ has strong effect on the currents on the NS superlattice, as shown in Fig.2 for a 2-NS-junction lattice. It is found that the probabilities of the reflected electron $|b_1|^2$, the outgoing electron $|a_{n+1}|^2$, the Andreev reflection $|c_1|^2$ and the hole reflection $|d_{n+1}|^2$ vary periodically with the the phase difference $\phi$, and so do the incident and outgoing currents. The total probability  is always conserved at different phase differences $|a_{n+1}|^2+|b_1|^2+|c_1|^2+|d_{n+1}|^2=2$. However, the incident and outgoing currents do not match each other at different phase differences(except two points). This result has never been reported in literatures and no one noticed that these two currents oscillate about the phase difference. Xing et al simply averaged the two currents from only in-phase incidences of an electron and a hole to balance the two currents\cite{Dong}. One of the present authors assumed in an earlier paper a charge accumulation on the interfaces to conserve the currents\cite{wliming}. They all neglected the effect of the phase difference on the currents. An important result of the present computation is that the incident and outgoing currents have almost the same average value for different $\phi$. This indicates that a great number of electrons and holes with random phase differences incident on the two leads of the NS junction superlattice will balance the two currents naturally. This provides a new but final mechanism for the balance of currents on the NS junction superlattice.

Using the above mechanism the bias voltage is scanned for the currents as shown in Fig.3. It is seen that the currents from in-phase incidences of an electron and a hole obviously deviate from each other above the superconducting  gap as reported by earlier works\cite{Wei, wliming}. After averaging over the phase difference the currents meet each other perfectly except small deviations at the three peaks of the currents. Therefore, this averaging process successfully recovers the balance of currents in a NS superlattice. The differential conductance for different interface barriers under this phase-averaging mechanism is shown in Fig.4(a). For smaller and smaller barrier potentials the differential conductance approaches the saturating value, i.e. fully conducting NS junctions. As the barrier increases the differential conductance shows stronger and stronger tunneling effect.  The first resonance peak takes place slightly higher than the superconducting gap. This provides a technique for measuring superconducting gaps. The other two resonance peaks occur at about $eV = 1.35\Delta, 1.8\Delta$. These resonances need experimental verification. For a comparison to Dong's work\cite{Dong}, where only in-phase incidence of electrons and holes is considered, Fig.4(b) shows the differential conductance using the phase averaging method and Dong's parameters.  For weak barriers the present result is quite different from Dong's result, but for strong barriers the present result approaches to Dong's result.

\begin{figure}
\includegraphics[width=4.cm,height=3.8cm]{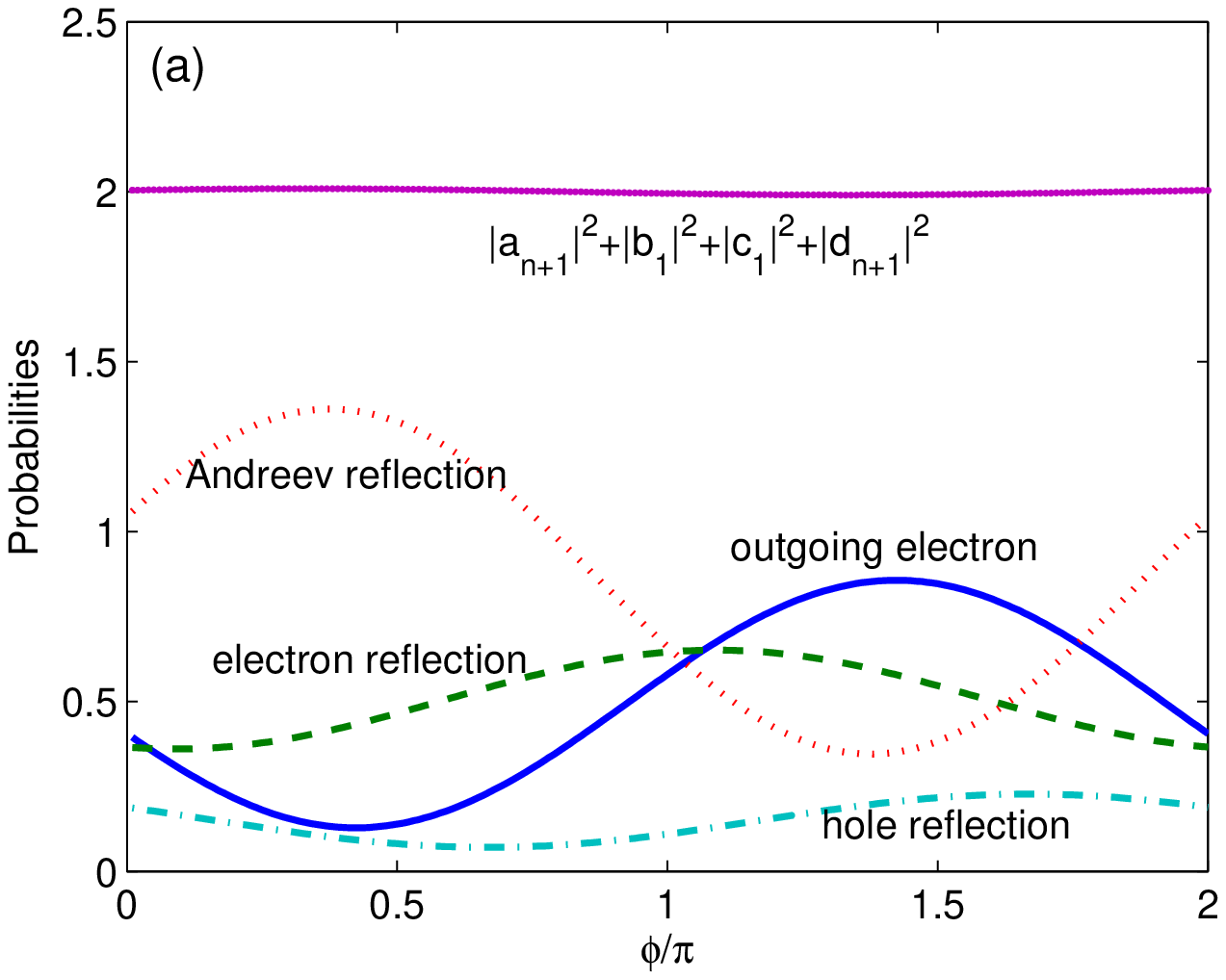}
\includegraphics[width=4.cm,height=3.8cm]{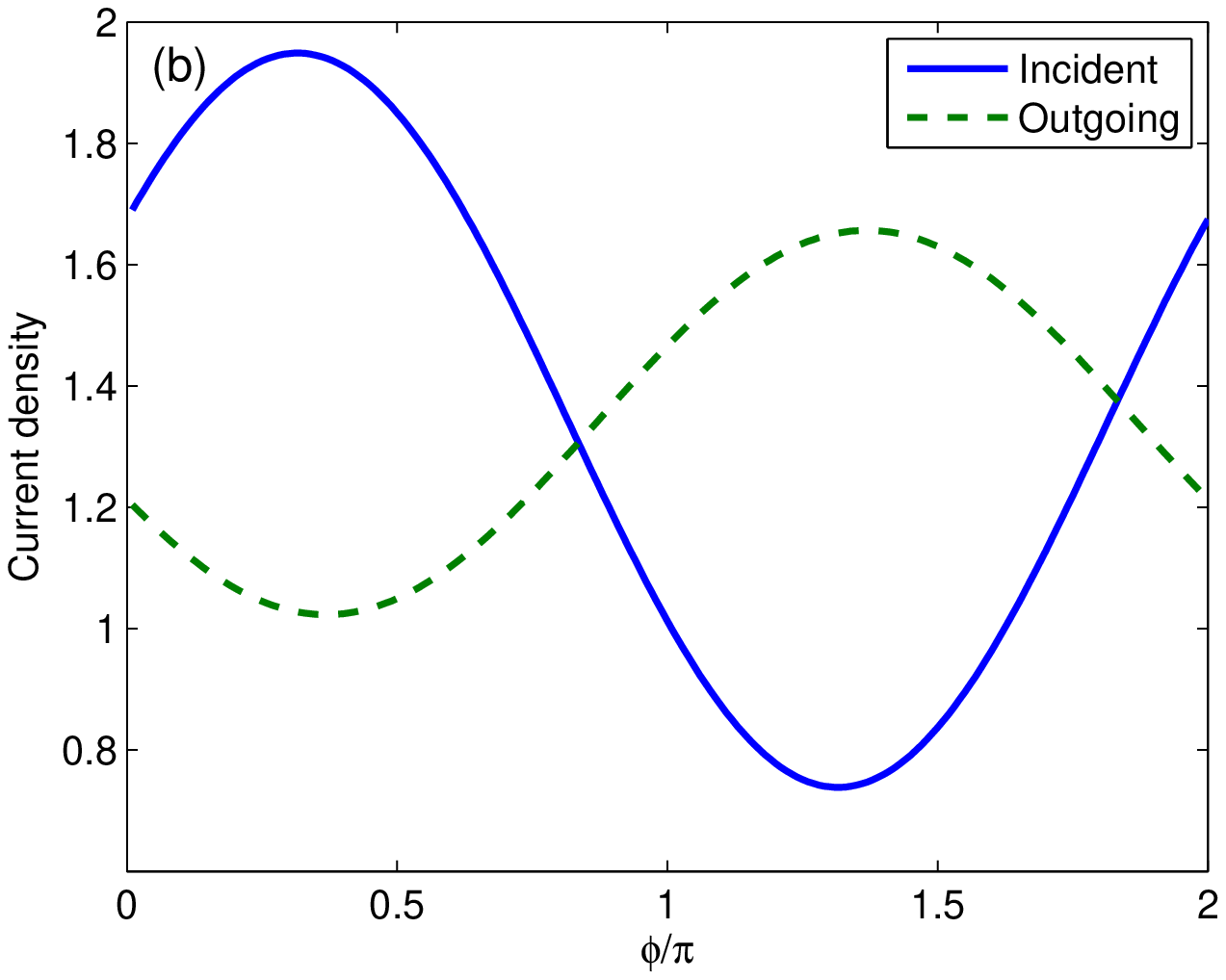}
\caption{(a) Probabilities of the reflected electron, the outgoing electron, the Andreev reflection, and the hole reflection in NSN junctions. The total probability of these terms is exactly equal to 2. (b) The incident current and the outgoing current $J_1$, $J_{n+1}$ at different phase  $\phi$. Parameters in the computation are set to be $\mu=0.5, \Delta = 0.01\mu, Z = 1, k_F L=1000, eV = 2\Delta$. }
\end{figure}

\begin{figure}
\includegraphics[width=4.cm,height=3.8cm]{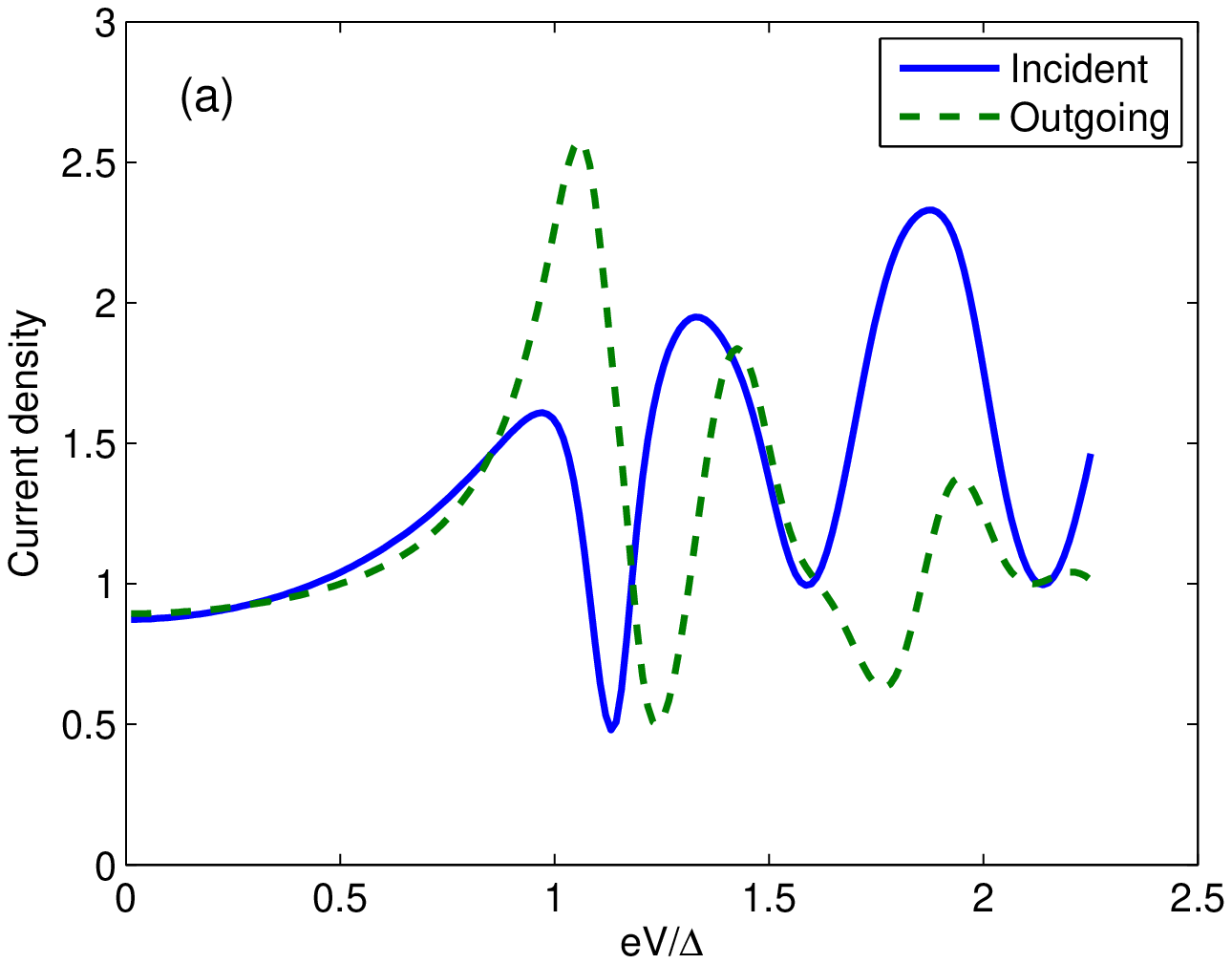}
\includegraphics[width=4.cm,height=3.8cm]{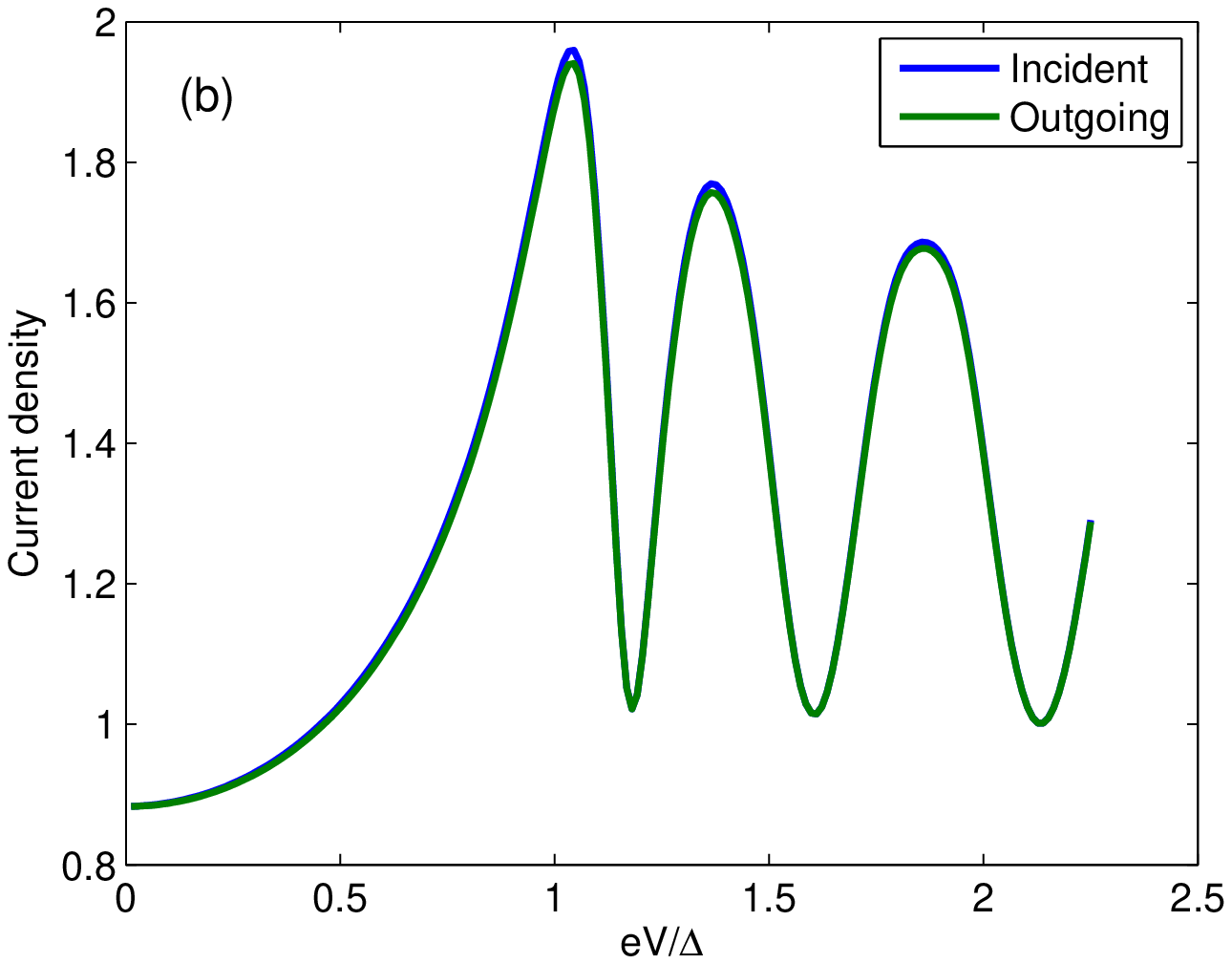}
\caption{(a) Incident and outgoing currents in NSN junctions with an in-phase incidence of an electron and a hole. (b) The currents averaged over the phase $\phi$. Parameters see Fig.2.}
\end{figure}

\begin{figure}
\includegraphics[width=4.cm,height=3.8cm]{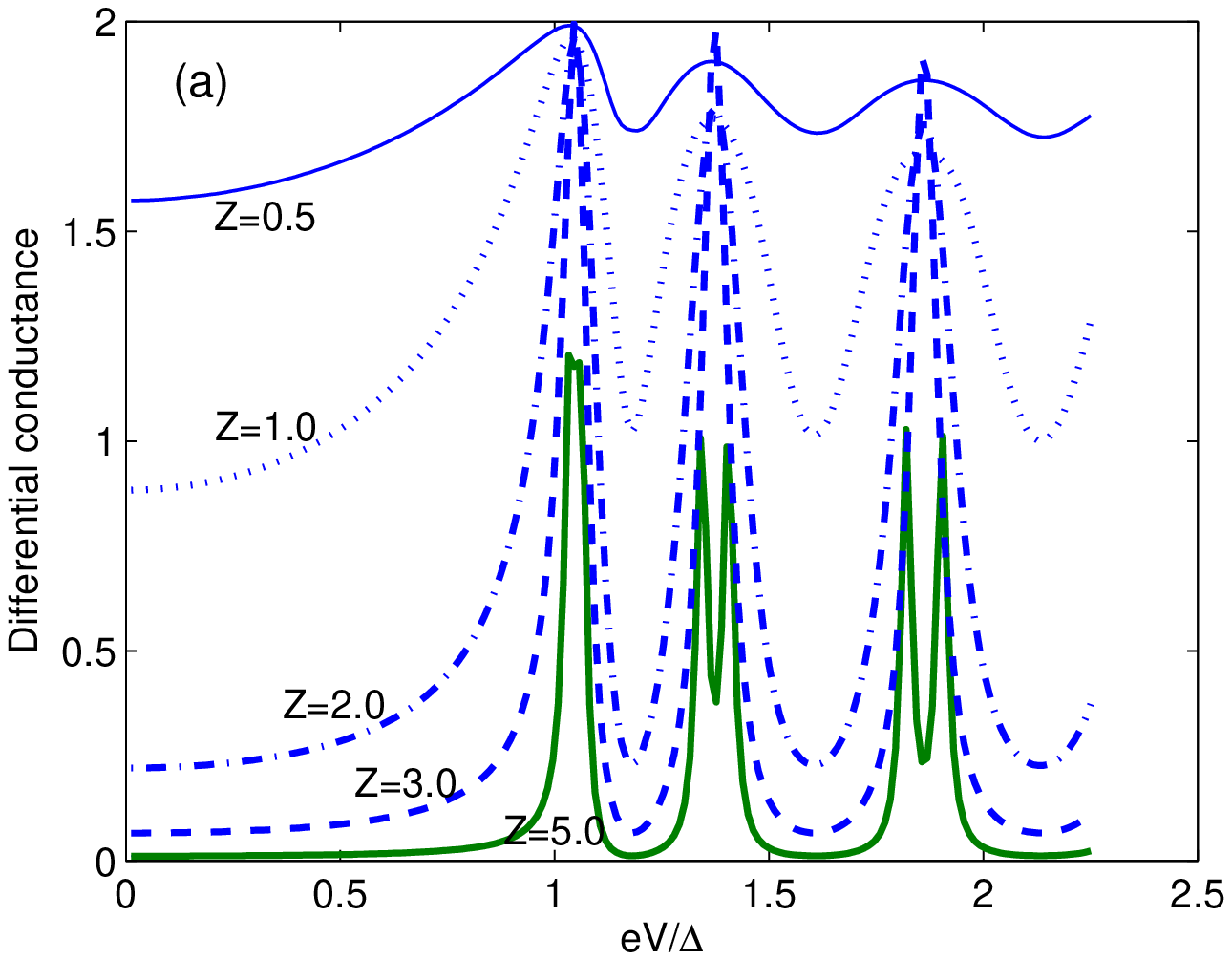}
\includegraphics[width=4.cm,height=3.8cm]{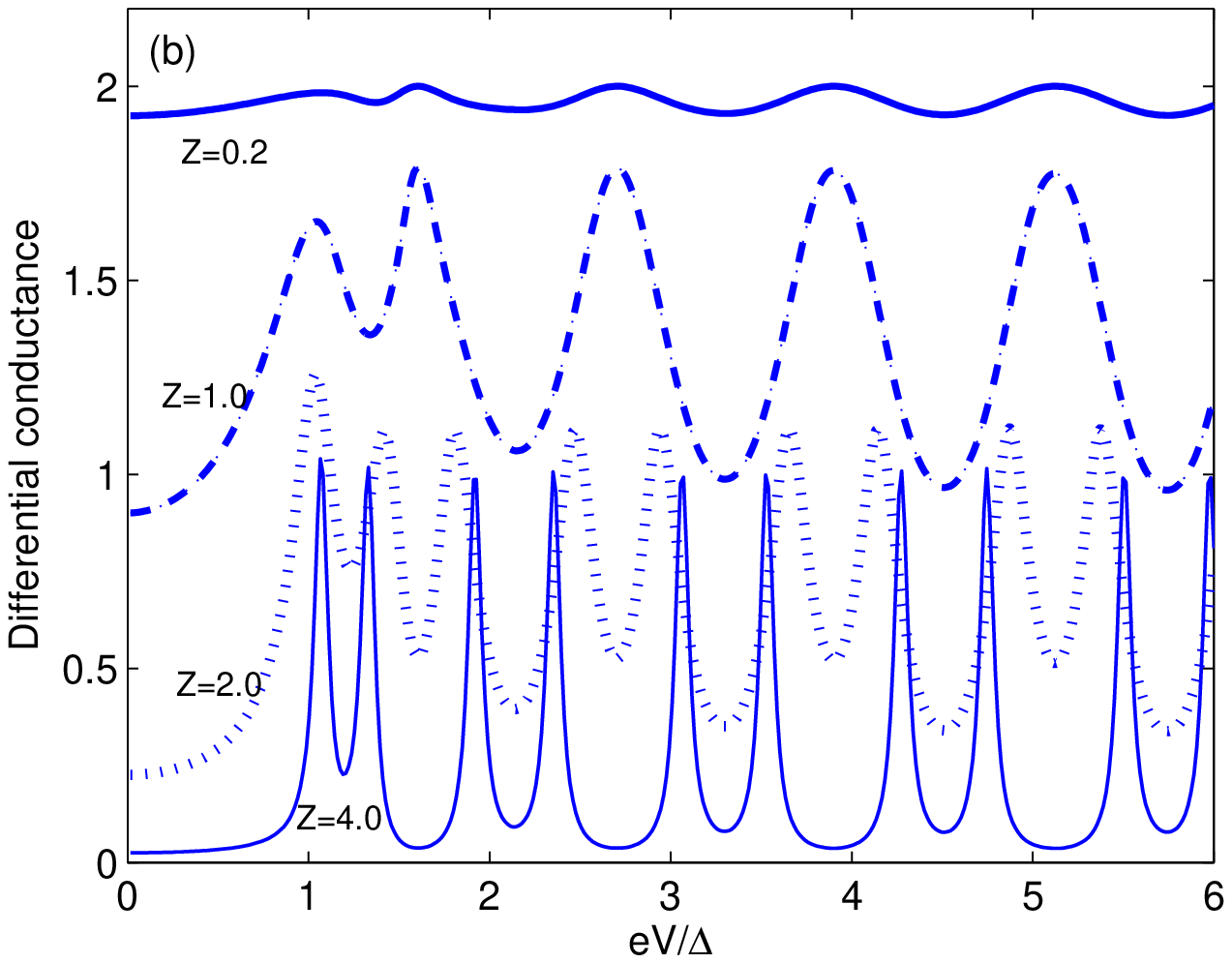}
\caption{Differential conductance for different interface barriers. (a) parameters see Fig.2; (b) A comparison to Fig.6 of Dong's work\cite{Dong} with parameters $\mu=0.5, \Delta = 0.001\mu, k_F L=5000$. }
\end{figure}

Next superlattices of 2,4,8 and 16 NS junctions are studied in the above phase averaging mechanism. The differential conductances of them with barriers $Z=0.5$ and $Z=4$ are shown in Fig.5(a,b). It is seen that in these two cases the NS junction superlattices have some sharp zero conductance resonances. Especially those of the superlattice of 16 junctions has a full width at half maximum of $0.012\Delta$, as seen in the inset of (b), which is equivalent to a voltage of about $0.5mV$. This effect may find applications such as  sensitive millivolt electronic switches, where a weak voltage increase induces a strong current response.

\begin{figure}\includegraphics[width=4.cm,height=3.8cm]{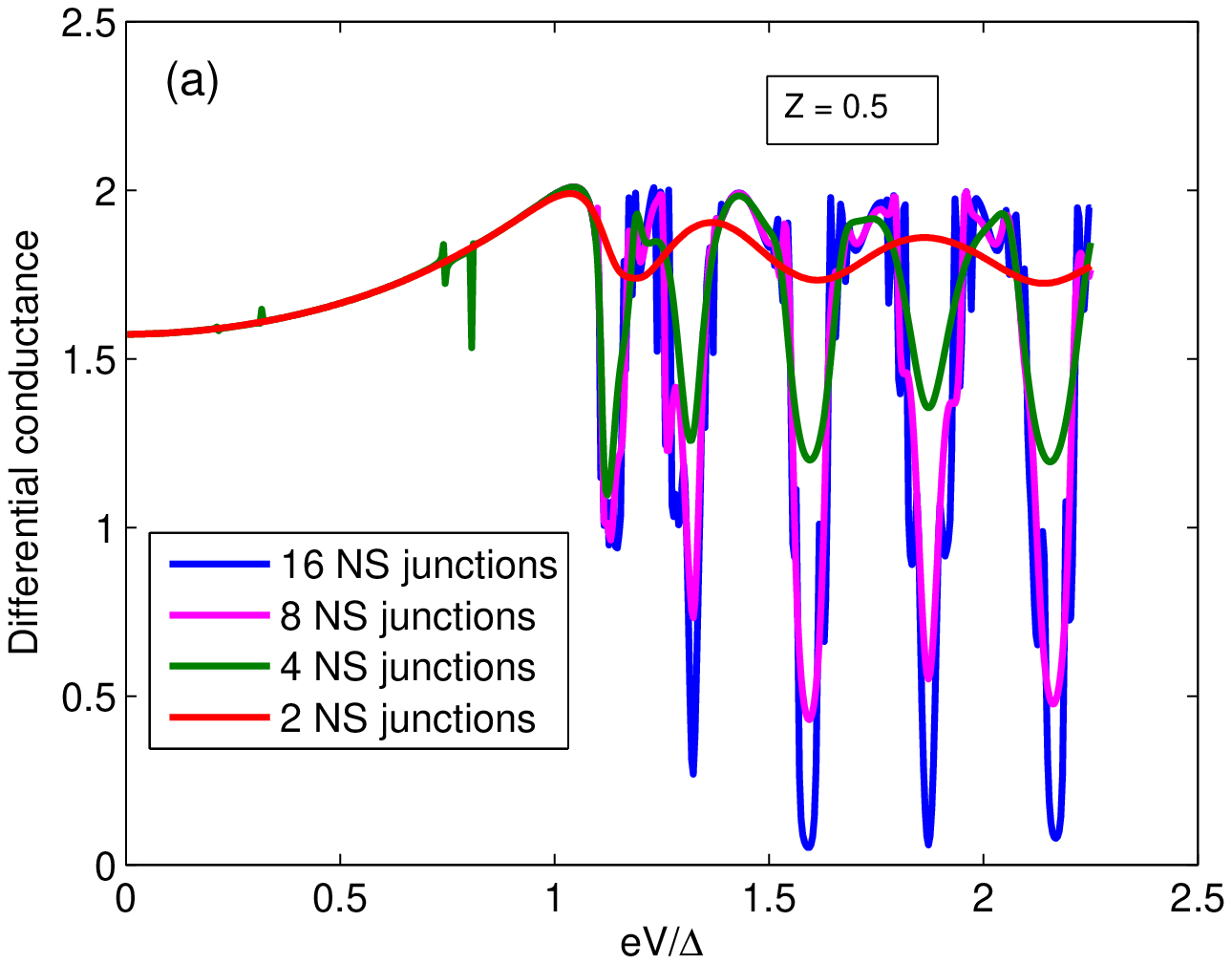}\includegraphics[width=4.cm,height=3.8cm]{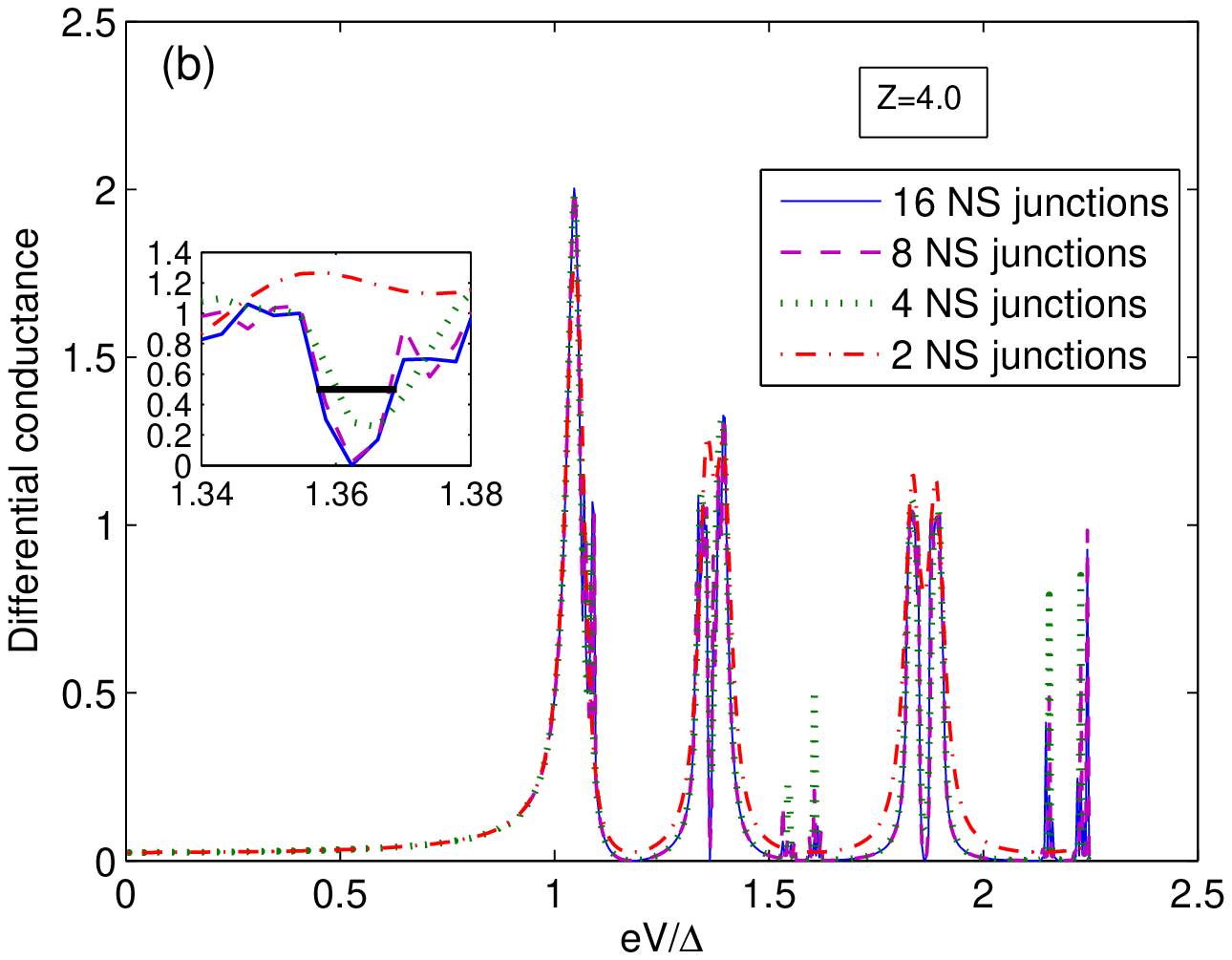}
\caption{Differential conductances of superlattices of 2,4,8, and 16 NS junctions with (a) Z = 0.5 and (b) Z=4.0. Other parameters see Fig.2.}
\end{figure}

Finally it should be pointed out that the first resonance peak of the differential conductance does not occur exictly at voltages corresponding to the energy gap of the SC at small values of $k_fL$. As shown in Fig.6 until $k_fL = 1200 $ the first resonance peak moves to the  gap position. This indicates that the measuring density of states using the technique of tunneling spectrum of metal-SC junctions requires a thickness of the SC layer above the minimum value.
\begin{figure}
\includegraphics[width=4.cm,height=3.8cm]{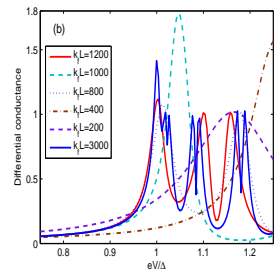}
\caption{Differential conductances of multiple NS junction lattices with different width $k_f L$ of superconducting layers. Other parameters see Fig.2.}
\end{figure}

In summary, the tunneling spectrum of an electron and a hole on a superlattice of NS junctions is computed using the BTK approach. It shows a more abundant structure compared to that on a single NS junction, such as the sharp resonance above the gap. In particular, the sharper the resonance is the more layers the superlattice has. This effect may find applications in the electronic industry in the future. We find  for the first time a  mechanism to balance the incident and outgoing currents on the superlattice  by averaging over the phase between the incident electron and the incident hole. Compared to other mechanisms such as the adjusting of the Fermi surface and the charge accumulation the present mechanism is much more natural and physical.

\section{acknowledgment}
 This work was supported by the National Natural Science Foundation of
China (Grant No. 10874049),  the State Key Program for Basic Research of China (No.
2007CB925204).

\end{document}